\documentstyle[preprint,aps,epsfig]{revtex}
\begin{document}
\preprint{\vbox{\baselineskip16pt
\hbox{KIAS-P98012}
\hbox{SNUTP 98-055}
}}
\title{
Excited heavy baryon spectrum 
in large $N_c$ heavy quark effective theory}
\author{
  Jong-Phil Lee$^a$\footnote{Email address:~jplee@phya.snu.ac.kr},
  Chun Liu$^b$\footnote{Email address:~liuc@ctp.snu.ac.kr}, and
  H.~S. Song$^a$\footnote{Email address:~hssong@physs.snu.ac.kr}}
\address{
  $^a$Department of Physics and Center for Theoretical Physics,\\
  Seoul National University, Seoul 151-742, Korea}
\address{
  $^b$Korea Institute for Advanced Study,\\
     207-43 Chungryangri-dong, Dongdaemun-ku, Seoul 130-012, Korea} 
\maketitle
\begin{abstract}
$L=1$ excited heavy baryon masses are analyzed by heavy quark and 
large $N_c$ expansions.
In heavy quark limit, mass is parameterized by $\bar{\Lambda}$ and it
is expanded further by spin-flavor breaking operators to the zeroth
order of $1/N_c$.  
Expanding coefficients will be fixed by more data on the excited baryons 
in the near future.
\end{abstract}
\pacs{14.20.G}
\section{Introduction}
In the near future, there will be large amount of data for orbitally excited
heavy baryons.  Understanding them will extend our ability in the application
of QCD \cite{Falk}.  
In this paper, we study their masses by using heavy quark and
large $N_c$ expansions.  
Heavy quark effective theory (HQET) \cite{Isgur} is a useful
method to deal with hadrons containing a single heavy quark.  
Many features of heavy mesons and heavy baryons have been analyzed 
by using HQET.  
In $m_Q\to\infty$ limit where $m_Q$ is the heavy quark mass, 
heavy quark spin decouples from the strong interaction and the hadron respects 
heavy quark spin-flavor symmetry (HQS).
To the order of $1/m_Q$, the breaking of the HQS occurs.  The excited heavy
baryon masses can be expressed as
\begin{equation}
M=m_Q+\bar{\Lambda}+\frac{c_1}{2m_Q} +\frac{c_2}{2m_Q}+O(\frac{1}{m_Q^2}),
\end{equation}
where the parameter $\bar{\Lambda}$ is independent of heavy quark spin and
flavor, and describes mainly the contributions of the light degrees of
freedom in the baryon. 
Here $c_1$ and $c_2$ are given by
\begin{eqnarray}
c_1&=&-\langle H_Q(v)|\bar{Q}_v(iD)^2Q_v|H_Q(v)\rangle\nonumber\\
c_2&=&-\frac{1}{2}Z_Q\langle H_Q(v)|\bar{Q}_v gG_{\mu\nu}
   \sigma^{\mu\nu}Q_v|H_Q(v)\rangle
\end{eqnarray}
where $H_Q(v)$ is the hadron state of velocity $v$ 
and $Z_Q$ is the renormalization factor.
The parameters $\bar{\Lambda}$, $c_1$, and $c_2$ should be determined 
by some nonperturbative methods.\par
At this stage, we use large $N_c$ expansion to analyze the parameter
$\bar{\Lambda}$.  Large $N_c$ QCD has been developed to study the
nonperturbative nature of hadrons \cite{Hooft}.  
In the large $N_c$ limit, baryons can
be treated as the bound state of infinite number of valence quarks. 
Witten described this system in the Hartree-Fock picture and gave $N_c$
counting rules for the meson-baryon scattering amplitudes \cite{Hooft}.
Dashen, Jenkins and Manohar (DJM) found that there is a light quark
spin-flavor symmetry for the ground state baryon sector in the large $N_c$
limit by deriving a set of consistency conditions 
(which was first derived by Gervais and Sakita \cite{Gervais})
for pion-baryon coupling
constants according to the $N_c$ counting rules \cite{DJM}.  
It was shown later that this symmetry can be observed 
in the Hartree-Fock picture \cite{Georgi,Luty}.  
Namely, the
s-wave states of low spin in the baryon multiplet are spin independent,
while the states with spin of order $N_c/2$ are considerably modified by
spin-spin and spin-orbit interactions.  Orbitally excited baryons have also
been studied [8-11].
As far as the baryon masses are concerned, 
Ref.\cite{Pirjol} derived the consistency conditions of the 
strong couplings of the excited baryons to pions analogous to that of 
Refs.\cite{Gervais,DJM}.
And Goity \cite{Goity}  made
the analysis for the excited light baryons in the Hartree-Fock picture.  
Further development was made in Ref.\cite{Lebed}.
We will adopt the Hartree-Fock picture to study $\bar{\Lambda}$.\par
One problem of our method is the validity of the large $N_c$ application.
It is actually $N_c-2$, which is $1$ in the real world, 
that will be taken as a large number, 
because heavy quark and the excited light quark are distinguished.
In the case of excited light baryons, the number which is taken to be 
large is $N_c-1$.
And it seems that the large $N_c$ approach describes the spectrum well 
\cite{Lebed}.
This leads us to generalize the method to the excited heavy baryons, 
considering that the theoretical framework is interesting and our knowledge
about non-perturbative HQET is poor.\par

This paper is organized as follows.
In Sec.II, excited heavy baryons are classified according to the large $N_c$
spin-flavor structure of light degrees of freedom.
In Sec.III, leading spin-flavor symmetry breaking operators 
involving spin-orbit, spin-isospin, and spin-orbit-isospin correlations
are introduced and their matrix elements are calculated in the basis given in
Sec.II.
The numerical results are obtained.
A summary is given in Sec.IV.
\section{Classification of the excited heavy baryons}
The quantum numbers which describe the hadrons are angular momentum $J$ and
isospin $I$.  
For the heavy hadrons, because of HQS, the total angular
momentum of the light degrees of freedom $J^l$ becomes a good quantum number.
In this case, the excited hadron spectrum shows the degeneracy of pair of
states which are related to each other by HQS.  
For the baryons, the light
degrees of freedom looks like a collection of $N_c-1$ light quarks with one of
them being excited.\par
Experimentally, some of the excited charmed baryons have been discovered
\cite{PDG}.
Among them, the lowest lying pair of states are
$\Lambda_{c1}(\frac{1}{2})^+$ and $\Lambda_{c1}(\frac{3}{2})^+$.
They are of isospin $I=0$.
In the constituent picture, the total spin of the light quarks $S^l$ is 
also zero \cite{Chiladze}.
This guides us to focus on the symmetric representation of the $N_c-1$ 
light quarks.\par
\begin{figure}
\vskip 2cm
\begin{center}
\epsfig{file=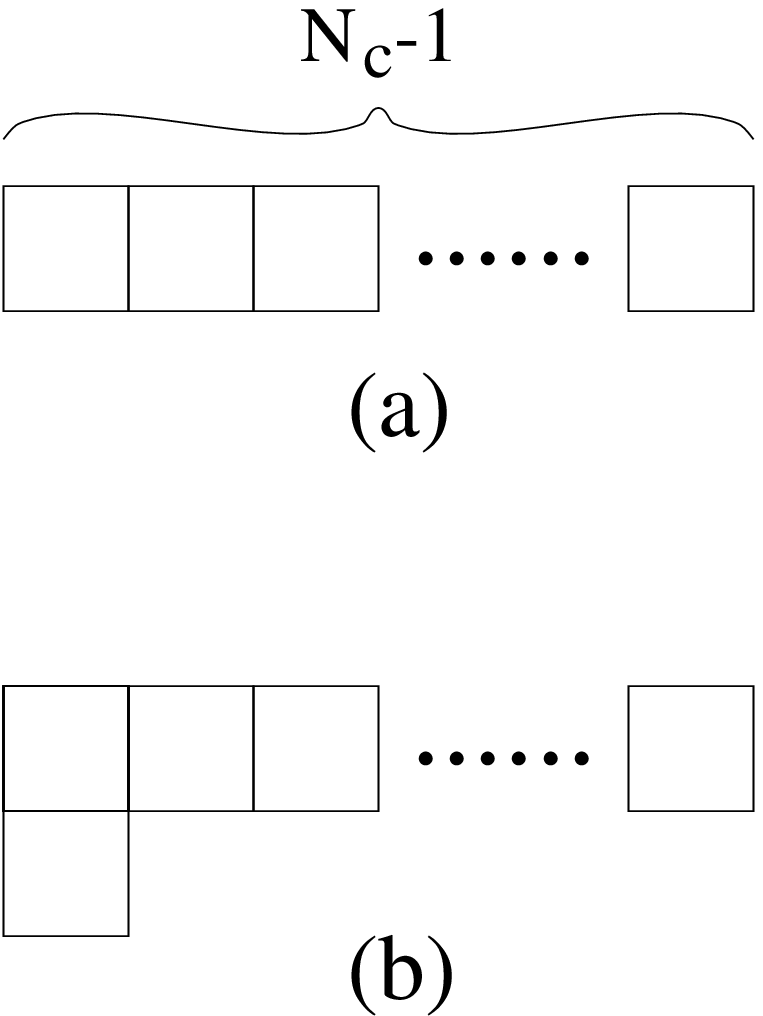, height=5cm}
\end{center}
\caption{}
\label{tableaux}
\end{figure}

Fig. \ref{tableaux}(a) shows the Young's tableaux of symmetric representation
of the $N_c-1$ quarks.
The spin-flavor decomposition is the same as that of the ground state heavy 
baryons \cite{DJM} with the rule $I=S^l$ for the non-strange baryons.
In this case, it should be noted that one of the light quarks is orbitally
excited with $l=1$.
In real world $N_c$ is fixed to be 3, so there are only two light quarks
in the heavy baryon, and one of which is excited.
The spin-flavor structure of these two light quarks is quite simple,
$(I,S^l)=(0,0)$ and $(1,1)$ by assuming $N_f=2$.
All the possible states of excited heavy baryons are listed in 
Table \ref{table1}.
The first two are $\Lambda_{c1}(\frac{1}{2})^+$ and 
$\Lambda_{c1}(\frac{3}{2})^+$, respectively.
For comparison, we also give out the mixed representation of the $N_c-1$
light quarks in Fig. \ref{tableaux}(b) and Table \ref{table2}.\par
It is convenient to introduce $K$-spin which is defined by
$\vec{K}=\vec{I}+\vec{S^l}$.
$K$ is a good quantum number whenever there is a light quark spin-flavor
symmetry which is true for baryons in the large $N_c$ limit.
The symmetric representation corresponds to $K=0$, and the mixed one to $K=1$.
From Tables \ref{table1} and \ref{table2}, we see that unlike the case of 
the excited light baryons \cite{Goity}, there is no mixing between the
states belonging to different $K$ for the case of excited heavy baryon.
Since there are only two Young boxes for light quarks, 
it is impossible to have the same $(I,S^l)$ pair with different $K$.\par
\section{Mass splittings}
With the classification of the excited heavy baryons described in the last
section, their spectrum, especially the mass splittings among them
will be studied by large $N_c$ method in the heavy quark limit.
If we really go to the large $N_c$ limit,
at the leading order ($N_c\Lambda_{\rm QCD}$),
we will get a trivial result, 
that is all the finitely excited heavy baryons have the same mass as that 
of the ground state heavy baryon.
This is simply because there are infinite number of light quarks,
which are not excited , in both excited and ground state baryons.
Compared to their contribution, the finitely excited quarks are negligible.
Such a conclusion is not useful practically, because in real world,
there is only one quark in the core.
Interesting point of recent approaches [5-7] is that the mass 
splittings due to the spin-flavor structure can be analyzed by large 
$N_c$ method.\par
Let us go to more details of large $N_c$ method.
One of the essential point of this method is the $N_c$ counting rules of the
relevant Feynman diagrams.
In the Hartree-Fock picture of the baryons, the $N_c$ counting rules require
us to include many-body interactions in the analysis, instead of including
only one- or two-body interactions. 
However, a large part of these interactions are spin-flavor irrelevant.
Namely, this part contributes in the order $N_c\Lambda_{\rm QCD}$ universally to
all the baryons with different spin-flavor structure.
The many-body Hamiltonians related to the spin-flavor structure which involve
orbital angular momentum $L$ give $O(1)$ contribution.
It is desirable to adopt a working assumption that they can be
treated perturbatively.\par
We use the following operators which are used in \cite{Goity} to 
analyze $\bar{\Lambda}$,
\begin{eqnarray}
H_{LS}&\propto&\hat{a}^{\dagger}~\vec{L}\cdot\vec{\sigma}~\hat{a}\nonumber\\
H_T&\propto&\frac{1}{N_c}G^{ia}G_{ia}\nonumber\\
H_1&\propto&\frac{1}{N_c}\hat{a}^{\dagger}~L^i\otimes\tau^a~\hat{a}~G_{ia}
\nonumber\\
H_2&\propto&\frac{1}{N_c}\hat{a}^{\dagger}\{L^i,L^j\}\otimes\sigma_i
\otimes\tau^a~\hat{a}~G_{ja}~.
\label{H_int}
\end{eqnarray}
The first one $H_{LS}$ is one-body Hamiltonian, while the others are 
two-body Hamiltonians.
$G_{ia}$ are the generators of  the spin-flavor symmetry group SU(4), given by
\begin{equation}
G^{ia}=\hat{a}^{\dagger}~\sigma^i\otimes\tau^a~\hat{a}
\end{equation}
with $\sigma^i$ and $\tau^a$ being the spin and isospin matrices, 
respectively.
Such structure gives coherent addition over $N_c-2$ core quarks.
The first $G^{ia}$ in $H_T$ acts on the excited quark,
the other $G_{ia}$'s on the $N_c-2$ unexcited light quarks, namely the
core quarks.
In our case, all the operators must be understood as the ones acting 
on the light degrees of freedom.
Note that the higher order many-body Hamiltonian which contains more factor
of $G_{ia}$ can be reduced to those given in Eq.(3) \cite{Goity}.
\par
The contributions to the baryon masses due to these Hamiltonians are 
obtained by calculating the baryonic matrix elements.
The matrix elements of these operators between the states of light quarks
which specify the states of excited heavy baryons are given as follows
analogous to Ref.\cite{Goity},
\begin{eqnarray}
&&\langle I_c=\frac{1}{2};~I ~I_3;~S^{l\prime}~S^{l\prime}_3, 
  ~l=1~m^{\prime}~|~H_T~|~
  I_c=\frac{1}{2};~I~I_3;~S^l~S^l_3,~l=1~m\rangle\nonumber\\
&=&2c_T\delta_{S^{l\prime},S^l}\delta_{S^{l\prime}_3,S^l_3}
  \delta_{m,m^\prime}(-1)^{1-S^l-I}
  \left\{\begin{array}{ccc}
  S^l&\frac{1}{2}&\frac{1}{2}\\
  1&\frac{1}{2}&\frac{1}{2}\end{array}\right\}
  \left\{\begin{array}{ccc}
  \frac{1}{2}&1&\frac{1}{2}\\
  \frac{1}{2}&I&\frac{1}{2}\end{array}\right\}~,\nonumber\\
\\
&&\langle I_c=\frac{1}{2};~I~I_3;~l=1,~S^{l\prime},~J^l~J^l_3~|~H_{LS}~|
  ~I_c=\frac{1}{2};~I~I_3;~l=1,~S^l,~J^l~J^l_3\rangle\nonumber\\
&=&c_{LS}(-1)^{S^l-S^{l\prime}}\sqrt{(2S^l+1)(2S^{l\prime}+1)}
  \sum_{j=\frac{1}{2},\frac{3}{2}}(2j+1)\{j(j+1)-2-3/4\}
  \left\{\begin{array}{ccc}
  \frac{1}{2}&\frac{1}{2}&S^l\\
  1&J^l&j\end{array}\right\}
  \left\{\begin{array}{ccc}
  \frac{1}{2}&\frac{1}{2}&S^{l\prime}\\
  1&J^l&j\end{array}\right\}~,\nonumber\\
\\
&&\langle I_c=\frac{1}{2};~I~I_3;~l=1,~S^{l\prime},~J^l~J^l_3~|~H_1~|
  ~I_c=\frac{1}{2};~I~I_3;~l=1,~S^l,~J^l~J^l_3\rangle\nonumber\\
&=&6c_1(-1)^{I-J^l+S^l-S^{l\prime}-1}
  \sqrt{(2S^l+1)(2S^{l\prime}+1)}
  \left\{\begin{array}{ccc}
  \frac{1}{2}&1&\frac{1}{2}\\
  \frac{1}{2}&I&\frac{1}{2}\end{array}\right\}
  \left\{\begin{array}{ccc}
  S^l&1&S^{l\prime}\\
  \frac{1}{2}&\frac{1}{2}&\frac{1}{2}\end{array}\right\}
  \left\{\begin{array}{ccc}
  1&1&1\\
  S^{l\prime}&J^l&S^l\end{array}\right\}~,\nonumber\\
\\
&&\langle I_c=\frac{1}{2};~I~I_3;~l=1,~S^{l\prime},~J^l~J^l_3~|~H_2~|
  ~I_c=\frac{1}{2};~I~I_3;~l=1,~S^l,~J^l~J^l_3\rangle\nonumber\\
&=&3c_2(-1)^{1+J^l+I+S^{l\prime}+2S^l}
  \sqrt{(2S^l+1)(2S^{l\prime}+1)}
  \left\{\begin{array}{ccc}
  \frac{1}{2}&1&\frac{1}{2}\\
  \frac{1}{2}&I&\frac{1}{2}\end{array}\right\}
  \left\{\begin{array}{ccc}
  2&1&1\\
  J^l&S^{l\prime}&S^l\end{array}\right\}
  \left\{\begin{array}{ccc}
  S^{l\prime}&S^l&2\\
  \frac{1}{2}&\frac{1}{2}&1\\
  \frac{1}{2}&\frac{1}{2}&1\end{array}\right\}~,
\end{eqnarray}
where $I_c$ is the isospin of the core quarks.
In real world ($N_c=3$), there is only one quark in the core ($N_c-2)$ 
so $I_c$ is always equal to $\frac{1}{2}$.\par
With these matrix elements, we can express the excited heavy baryon mass $M$
up to the zeroth order of $1/m_Q$ and $1/N_c$ as follows.
\begin{eqnarray}
M&=&m_Q+\bar{\Lambda}~,\nonumber\\
\bar{\Lambda}&=&\Lambda_0 +\langle H_{LS}\rangle+\langle H_T\rangle+
  \sum_{i=1}^{2}\langle H_i\rangle~,
\end{eqnarray}
where $c_{LS}$, $c_T$, and $c_i$'s are the coefficients which will be 
determined by experiments and $\Lambda_0$ is the leading contribution to 
$\bar{\Lambda}$ which preserves the spin-flavor symmetry.\par
The numerical results are also given on the right hand side of 
Table \ref{table1} and \ref{table2}.
Current data are 
$\Lambda_{c1}^+(J=\frac{1}{2},\frac{3}{2})=2593.9\pm0.8,~2626.6\pm0.8
~({\rm MeV})$ \cite{PDG}.
This results in 
$-\frac{1}{6}c_T\simeq 0.1$~GeV by taking $m_c=1.5$~GeV and 
$\Lambda_0=N_c\Lambda_{\rm QCD}=1.0$~GeV.
More experimental data are needed to fix the unknown coefficients.
In the work for the light excited baryons \cite{Goity}, 
$\langle H_T\rangle$ yields the same value for a given tower so it can be
absorbed into the leading contribution. 
In our case, as can be seen in Table \ref{table1}, 
$\langle H_T\rangle$ gets different
values in symmetric representation while in mixed representation
it gives the same value for all the states of the tower.
Comparing with the mixed representation, mass splittings of the 
symmetric states due to $H_T$ and $H_i$ are opposite in sign and three times
smaller except for $(J^l,S^l)=(1,0)$.
This is due to the effect of different isospin.
In the future, these coefficients can be fitted after obtaining the three
masses of the states with quantum numbers
$(J^l,S^l)=(0,1),~(1,1)$ and $(2,1)$ in the symmetric representation.
It would be a check for the validity of our method, if the results for
$c_{LS}$, $c_1$ and $c_2$ are in the reasonable range as that for $c_T$.
\section{Summary}
We have analyzed the mass splittings of orbitally excited heavy baryons in terms
of $1/m_Q$ and $1/N_c$ expansions.
In heavy quark limit, heavy quark spin decouples and the baryon is 
described by light degrees of freedom.
At the zeroth order of $1/N_c$, the light quark spin-flavor symmetry breaking
effects which involve one 1-body operator and three 2-body operators have
been calculated,
and they are parameterized by several coefficients which need more data on
excited charmed baryons to be fitted.
\begin{center}
{\large\bf Acknowledgments}\\[10mm]
\end{center}\par
This work was supported in part by KOSEF through SRC programm.
Part of it was finished when one of us (C.L.) was visiting the University of
Notre Dame.
C.L. would like to thank Ikaros Bigi for hospitality.

\newpage
%
%
%
%
%
\begin{center}{\large\bf FIGURE CAPTIONS}\end{center}
\noindent
Fig.~1
Young's tableaux for (a) symmetric and (b) mixed representation of $N_c-1$
light quarks.\\
\vskip .3cm
\begin{center}{\large\bf TABLE CAPTIONS}\end{center}
\noindent
Table 1.
Excited heavy baryon states of the symmetric representation of $N_c-1$
light quarks.\\
\vskip .3cm
\noindent
Table 2.
Excited heavy baryon states of the mixed representation of $N_c-1$
light quarks.\\
\vskip .3cm
\newpage
%
%
\begin{table}
\vskip 5mm
\begin{center}
\begin{tabular}{ccc}
$(J,I)$ & $(J^l,S^l)$ & 
 $\langle H_I\rangle =\langle H_{LS}\rangle +\langle H_T\rangle
 +\sum_{i=1}^{2}\langle H_i\rangle$\\\hline
$(1/2,0)$ & $(1,0)$ & $-\frac{1}{6}c_T$\\
$(3/2,0)$ & $(1,0)$ & $-\frac{1}{6}c_T$\\
$(1/2,1)$ & $(0,1)$ &
$-2c_{LS}-\frac{1}{54}c_T+\frac{1}{9}c_1-\frac{1}{6}c_2$\\
$(1/2,1)$ & $(1,1)$ & 
$-c_{LS}-\frac{1}{54}c_T+\frac{1}{18}c_1+\frac{1}{12}c_2$\\
$(3/2,1)$ & $(1,1)$ &
$-c_{LS}-\frac{1}{54}c_T+\frac{1}{18}c_1+\frac{1}{12}c_2$\\
$(3/2,1)$ & $(2,1)$ &
$9c_{LS}-\frac{1}{54}c_T-\frac{1}{18}c_1-\frac{1}{60}c_2$\\
$(5/2,1)$ & $(2,1)$ &
$9c_{LS}-\frac{1}{54}c_T-\frac{1}{18}c_1-\frac{1}{60}c_2$
\end{tabular}
\end{center}
\caption{}
\label{table1}
\end{table}
\begin{table}
\vskip 5mm
\begin{center}
\begin{tabular}{ccc}
$(J,I)$ & $(J^l,S^l)$ &
 $\langle H_I\rangle=\langle H_{LS}\rangle+\langle H_T\rangle
 +\sum_{i=1}^{2}\langle H_i\rangle$\\\hline
$(1/2,0)$ & $(0,1)$ &
$-2c_{LS}+\frac{1}{18}c_T-\frac{1}{3}c_1+\frac{1}{2}c_2$\\
$(1/2,0)$ & $(1,1)$ &
$-c_{LS}+\frac{1}{18}c_T-\frac{1}{6}c_1-\frac{1}{4}c_2$\\
$(3/2,0)$ & $(1,1)$ &
$-c_{LS}+\frac{1}{18}c_T-\frac{1}{6}c_1-\frac{1}{4}c_2$\\
$(3/2,0)$ & $(2,1)$ &
$9c_{LS}+\frac{1}{18}c_T+\frac{1}{6}c_1+\frac{1}{20}c_2$\\
$(5/2,0)$ & $(2,1)$ &
$9c_{LS}+\frac{1}{18}c_T+\frac{1}{6}c_1+\frac{1}{20}c_2$\\
$(1/2,1)$ & $(1,0)$ & $\frac{1}{18}c_T$\\
$(3/2,1)$ & $(1,0)$ & $\frac{1}{18}c_T$
\end{tabular}
\end{center}
\caption{}
\label{table2}
\end{table}
\end{document}